\documentclass[prl, superscriptaddress, twocolumn, showpacs,
nofootinbib]{revtex4}

\usepackage[]{graphicx}
\usepackage{amsmath}
\usepackage{amsfonts}
\usepackage{color}
\usepackage{hyperref}
\usepackage{bm}
\usepackage{graphicx}
\usepackage{amsbsy}

\def\ket#1{| #1 \rangle}

\def\kb#1#2{|#1\rangle\!\langle #2 |}

\newcommand{\id}{\mathbb{I}}

\date{\today}

\begin{document}

\title{A Single Photon Source With Linear Optics and Squeezed States}

\author{Casey R. Myers}
\affiliation{Institute for Quantum Computing, University of
Waterloo, ON, N2L 3G1, Canada}
\author{Marie Ericsson}
\affiliation{Institute for Quantum Computing, University of
Waterloo, ON, N2L 3G1, Canada}\affiliation{Centre for Quantum
Computation, DAMTP, University of Cambridge, Wilberforce Road,
Cambridge CB3 0WA U.K.}
\author{Raymond Laflamme}
\affiliation{Institute for Quantum Computing, University of
Waterloo, ON, N2L 3G1, Canada} \affiliation{Perimeter Institute
for Theoretical Physics, 35 King Street N., Waterloo, ON, N2J 2W9,
Canada}

\begin{abstract}
  
  Quantum information processing using photons has recently been
  stimulated by the suggestion to use linear optics, single photon
  sources and detectors. The recent work by Knill has also shown that
  errors in photon detectors leads to a high error rate threshold
  (around $29\%$).  An important missing element are good single
  photon sources.  In this paper we show how to make a single photon
  source using squeezed states, linear optics and conditional
  measurement.  We use degenerate squeezed vacuum states, in
  contrast to the normal non-degenerate squeezed vacuum states used for
  single photon production. We show that we can get a photon with certainty
  when detectors click appropriately, the last event happening up to
  around $25\%$ of the time.  We also show the robustness of this
  method with respect to a variety of potential imperfections.

\end{abstract}

\maketitle


Quantum mechanics provides a new way to manipulate and encode
information that has no classical counterpart. This has lead to new
algorithms, cryptographic protocols, higher precision measuring
techniques and more.  Many proposals have been put forward to build
devices that will be able to harness the quantum world and turn the
theoretical advantage of using quantum mechanics into a practical
one \cite{areview}.  Quantum optics has demonstrated a high control of
the quantum world because photons interact weakly with each other and
thus have long decoherence times. Demonstration of quantum cryptography abound and have even lead to prototypes which are presently reaching the market. However these devices have severe limitation in the distance over which they can be used. To go beyond these
first prototypes by extending the maximum distance they can be deployed
or to increase their reliability, better control of the quantum systems
must be achieved in order to be able to implement quantum repeaters
and error correction schemes.  Such schemes require quantum gates and
interaction between photons.

Photons interact in media through what is called the Kerr
effect, but typically this interaction is much too weak for quantum
information purposes. One of the first experimental demonstrations
of quantum gates used cavity QED mediated interaction of photons \cite{Turchette}.
It was a nice proof of principle but it was however quickly
realized that it would be difficult to scale beyond a few gates.

Recently it has been proposed to use single photon sources, detectors
and linear optics \cite{KLM} for quantum information processing.  This
suggestion uses projection of the quantum states and feedback to
simulate efficiently the quantum circuit model. Progress in
LOQC has been made in two fronts. Proof-of-principle experiments,
demonstrating that the fundamental elements have been
realized \cite{Obrien, Pittman, Sanaka, Fattal}. These show that today we have sufficient
control to manipulate small quantum systems. The second avenue of
progress has been to assess the constraint in precision
of the elements required for LOQC.  Linear optical
elements, beam splitters and phase shifters, can be made with high
accuracy and are very reliable. These could without too much
difficulty reach the threshold accuracy for generic error
\cite{threshold} in order to compute reliably using quantum error
correction.  It was shown in \cite{knill:qc2003}
that, if only errors in detectors were present, an accuracy
threshold around $70\%$ was sufficient. A similar calculation with a
different set of assumptions showed an error threshold around $2\%$ \cite{Silva}.
Finally, although a lot of effort has been
put into making single photon sources, none are yet of a quality
sufficient for making the linear optics proposal scalable. In this
letter we propose a new way to devise such a source based on the idea of
linear optics.

The purpose of this paper is to propose a single photon source
that can be improved dramatically using the ideas of LOQC and
error correction, i.e. we are asking if it is possible to use a
set of imperfect sources, make them interfere through beam
splitters and phase shifters, detect some of the output, and
improve the amplitude of observing a single photon.  It turns out
that it seems impossible if the input is in a mixture of the state
$\ket{0}$ and $\ket{1}$ \cite{Imamoglu, Berry04:1, Berry04:2}. On the
other hand, if the source is coherent it is however possible.

\begin{figure}[ht!]
\begin{center}
\includegraphics[scale=0.4]{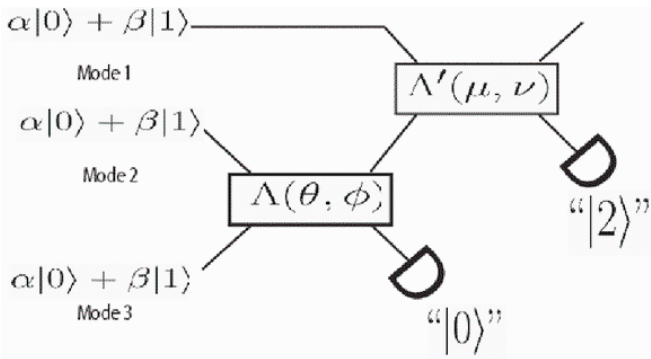}
\end{center}
\caption{\footnotesize Show the system to get a single photon from three identical known copies of the state $\alpha\ket{0}+\beta\ket{1}$.}
\label{singlphotdiag}
\end{figure}

If three identical copies of the state $\alpha \ket{0} +\beta\ket{1}$
are put onto the two beam splitter system \cite{Berry04:2} shown in
figure (\ref{singlphotdiag}), we are able to condition off of certain
detections to have, with certainty, the state $\ket{1}$ in the output
mode. This input state can be written as $\Bigl((\alpha+\beta
\hat{a}^{\dagger} )\ket{0}\Bigr)^{\otimes 3}$. We use the beam
splitter transformation \cite{KLM}:
$\hat{a}_{l}^{\dagger}\rightarrow\sum_{k}\Lambda_{kl}\hat{a}_{k}^{\dagger}$,
where
\begin{equation}
\label{Lambda}
\bm{\Lambda} = \left[ \begin{array}{*{20}c} \cos{\theta}
& -e^{i\phi}\sin {\theta} \\ e^{-i\phi}\sin\theta & \cos\theta  \\
\end{array} \right]
\end{equation}
If we condition off detecting ``2'' photons in mode 2 and ``0''
photons in mode 3 the state in mode 1 is (un-normalised)
\begin{eqnarray}
\label{twobssytem}
\beta^{2}\Lambda_{22}' [\alpha\bigl( \Lambda_{23}\Lambda_{21}' +\Lambda_{22}(\Lambda_{21}'+\Lambda_{23}\Lambda_{22}') \bigr)\ket{0}\nonumber\\ +  \beta\Lambda_{22}\Lambda_{23}\bigl(2\Lambda_{12}'\Lambda_{21}' + \Lambda_{11}'\Lambda_{22}'   \bigr)\ket{1}]
\end{eqnarray}
By choosing the
angles of both beam splitters to be a function of the beam splitter
phases $\phi$ and $\nu$ (see figure (\ref{singlphotdiag})), we can set the coefficient in front of
$\ket{0}$ in (\ref{twobssytem}) to 0 for: $\theta=\arctan \bigl(
\frac{\sin(\nu+\phi)}{\sin(\nu)} \bigr)$, $\mu=-\arctan \bigl(
\frac{\sin (\nu + \phi)\sin(\nu)}{\sin (\phi)
  \sqrt{\sin^{2}(\nu)+\sin^{2}(\phi+\nu)}}\bigr)$. The maximum
probability of a ``2'' ``0'' detection is $\frac{16|\beta|^{3}}{81}$
(but note that when we have these detections we insure having a single photon).

The difficulty is reduced to make a state of the form 
$\alpha\ket{0}+\beta\ket{1}$. 
Instead of such a state let's concentrate on a
squeezed coherent state. This has the form
$\ket{\xi,\alpha}=\frac{\exp[-\frac{1}{2}(|\alpha|^{2}-\alpha^{2}\tanh
  r)]}{\sqrt{\cosh
    r}}\sum_{n=0}^{\infty}\frac{1}{\sqrt{n!}}\Biggl(\sqrt{\frac{\sinh
    r}{2\cosh r}} \Biggr)^{n} \times
H_{n}\Bigl(\frac{\alpha}{\sqrt{2\cosh r\sinh r}}\Bigr)\ket{n}$ where
$\xi=r e^{i \varphi}$. As shown in \cite{Matsuoka}, we can set the 2
photon term to zero by setting the 2nd Hermite polynomial $H_{2}(x)$
to 0. Using this method we could generate states with a probability of
production equal to $1.2\%$ and a single photon content as high as
$96.5\%$ at the highest yet attainable squeezing ($r=0.36$
\cite{Wenger}). As we increase the squeezing (with the displacement
length changing according to $\alpha=\sqrt{\frac{\sinh (2r)}{2}}$ so
that $H_{2}(x)=0$), the probability of production increases until
$r=0.81$, after which it begins to drop towards 0. Also, as we
increase the squeezing, the single photon content percentage
asymptotes to a value of $82\%$. The problem with this scheme is that
there is no way to rid ourselves of higher order photon terms.

The above analysis motivates us to look at squeezed states as inputs into linear optical systems to try and better current single photon sources. Squeezed states are considered to be Gaussian states, the simplest type of Gaussian state is the coherent state, having many similarities to a
classical description of a field. A coherent state is defined by \cite{Walls} $\ket{\alpha}=\exp{(\alpha \hat{a}^{\dagger}-\alpha^{*}\hat{a})}\ket{0}$, where $a^{\dagger}\ket{n}=\sqrt{n+1}\ket{n+1}$. The light
from a stabilized laser is a typical example of a coherent state \cite{Ralph}. The Wigner function of a coherent state is given by \cite{Walls} $W(x_{1}', x_{2}')=\frac{2}{\pi}\exp\Bigl( -\frac{1}{2}(x_{1}'^{2}+x_{2}'^{2})\Bigr)$, where $x_{i}'=x_{i}-X_{i}$ and $X_{1}$ and $X_{2}$ play the role of non-commuting electric field quadrature observables:  $\hat{a}=\frac{1}{2}\bigl( X_{1}+i X_{2}\bigr)$. As can be seen a coherent state is a Gaussian in phase space centred at $(X_{1}, X_{2})$ with an equal distribution in all directions, that is a coherent state has minimum uncertainty in both the position and the momentum quadrature. 

Squeezed states also have a minimum Heisenberg uncertainty but with the uncertainty in one direction reduced at the expense of the uncertainty in the other direction. A squeezed vacuum is defined by \begin{eqnarray}
\label{squeezedef}
\ket{\xi}&=\exp\Bigl(\frac{\xi}{2}(\hat{a}^{\dagger})^{2}-\frac{\xi^{*}}{2}(\hat{a})^{2}\Bigr)\ket{0}
\end{eqnarray}
where $\xi=r\exp{(i\varphi)}$ governs the amount and orientation of
squeezing. The Wigner function for a squeezed state, squeezed along
one of the quadrature axes, is given by \cite{Walls} $W(x_{1}',
x_{2}')=\frac{2}{\pi}\exp\Bigl(
-\frac{1}{2}(x_{1}'^{2}e^{-2r}+x_{2}'^{2}e^{2r})\Bigr)$.  This is also
a Gaussian state centred at $(X_{1}, X_{2})$ but instead of a circle
in phase space, as with a coherent state, it is an ellipse, with minor
and major axes given by $\exp(r)$ and $\exp(-r)$, respectively.

Squeezed states are produced when coherent light interacts with a
nonlinear medium, such as in degenerate parametric down conversion
\cite{Wu86}. Squeezing can be achieved with either cw \cite{Polzik} or
pulsed light \cite{Slusher}, and this normally dictates what type of
measurement is allowable. It is only the latter type of squeezing,
that is pulsed squeezing, that allows the use of photon counting
detectors. It is for this reason that we consider pulsed squeezing in
this letter. To date the best cw squeezing achieved is 7.2dB \cite{Schneider}. This
corresponds to $r=0.83$. In contrast the best pulsed squeezing
recorded is 3.1dB \cite{Wenger}, corresponding to $r=0.36$.  This is
approximately an ellipse in phase space with a minor axis
$\frac{1}{2}$ the size of the major axis.

\begin{figure}[ht]
\includegraphics[width=8.5cm,height=3.5cm]{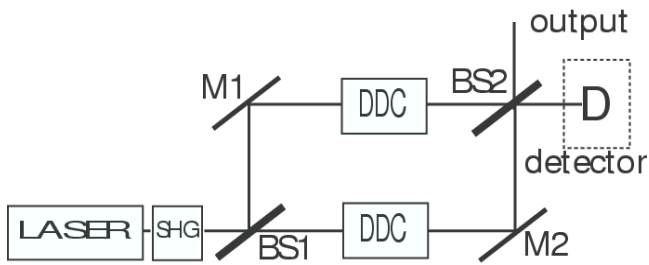}
\caption{\footnotesize Setup to allow two squeezed vacuum states onto a beam splitter. SGH = second harmonic generation, M = mirror, BS = beam splitter and DDC = degenerate down conversion. BS1 is a symmetric 50:50 beam splitter, as is BS2 ($\theta=\frac{\pi}{4}, \phi=\frac{\pi}{2}$). The dotted box indicates an inefficient detector.} 
\label{1BSsysFig}
\end{figure}

The squeezed vacuum state given in Eqn.(\ref{squeezedef}) is equal to \cite{Barnett, Vogel}
\begin{eqnarray}
\label{Soperator}
\ket{\zeta}=\frac{1}{\sqrt{\cosh{r}}}\exp[-\frac{1}{2}\exp(i\varphi)\tanh(r)(\hat{a}^{\dagger})^{2}]\ket{0}
\end{eqnarray}
If we put two of these states ($\ket{\xi_{1}}\otimes\ket{\xi_{2}}$) onto a beam splitter BS2, as shown in figure (\ref{1BSsysFig}), the term in the exponent becomes 
\begin{eqnarray}
\label{VacMiddle}
\Bigl(\lambda_{1}\Lambda_{11}^{2}+\lambda_{2}\Lambda_{12}^{2}\Bigr)(\hat{a}_{1}^{\dagger})^{2}&+&2\Bigl(\lambda_{1}\Lambda_{11}\Lambda_{21}+\lambda_{2}\Lambda_{12}\Lambda_{22} \Bigr)\hat{a}_{1}^{\dagger}\hat{a}_{2}^{\dagger}\nonumber\\
&+&\Bigl(\lambda_{1}\Lambda_{21}^{2}+\lambda_{2}\Lambda_{22}^{2}\Bigr)(\hat{a}_{2}^{\dagger})^{2}
\end{eqnarray}
where $\xi_{j}=r_{j}\exp(i\varphi_{j})$ and $\lambda_{j}=-\frac{1}{2}\exp(i\varphi_{j})\tanh(r_{j})$, $j\in\{1,2\}$.

If we detect off measuring a ``1''  in mode 2 this becomes: 
\begin{eqnarray}
\label{sqvacdet}
\frac{2}{\sqrt{\cosh(r_{1})\cosh(r_{2})}}\Bigl(\lambda_{1}\Lambda_{11}\Lambda_{21}+\lambda_{2}\Lambda_{12}\Lambda_{22} \Bigr)\hat{a}_{1}^{\dagger}\nonumber\\
\times\exp[\Bigl(\lambda_{1}\Lambda_{11}^{2}+\lambda_{2}\Lambda_{12}^{2}\Bigr)(\hat{a}_{1}^{\dagger})^{2}]\ket{00}\label{VacDet}
\end{eqnarray}
In order to have a perfect single photon source we need $\lambda_{1}\Lambda_{11}^{2}+\lambda_{2}\Lambda_{12}^{2}=0$. If we assume identical squeezed vacuum states such that $\lambda_{1}=\lambda_{2}$ and a symmetric beam splitter ($\theta=\frac{\pi}{4}, \phi=\frac{\pi}{2}$) this gives a single photon. Putting these numbers into Eqn.(\ref{VacMiddle}) gives the equivalent of a two mode squeezed vacuum \cite{Barnett}. Producing a two mode squeezed vacuum this way is advantageous for a few reasons, one being that the intensity of the pump beams need not be as high as that for non-degenerate down conversion. 

Before any detections, the state after BS2 in figure (\ref{1BSsysFig}) is
\begin{eqnarray}
\frac{1}{\cosh(r)}\exp [i \exp(i \varphi)\tanh(r) \hat{a}_{1}^{\dagger}\hat{a}_{2}^{\dagger}]\ket{00}\label{perfectstate1}
\end{eqnarray}
From (\ref{perfectstate1}) we can see that the probability of detecting $n$ photons in mode 2 is given by 
\begin{eqnarray}
\text{Prob.}=\frac{\tanh^{2n}(r)}{\cosh^{2}(r)}
\end{eqnarray}
The maximum probability to observe a single photon is $25\%$ at a squeezing of $r_{\text{max}}=0.88137$. With todays technology a squeezing of $r=0.36$ is possible \cite{Wenger}, corresponding to a single photon probability of $10.5\%$.

One problem that our scheme will face is mode mismatch on BS2. However it is possible to take this into account, as shown in \cite{Banaszek}. 


Next we consider how robust our proposed single photon source is
against an inefficient detector. We model our inefficient detector with a beam
splitter of reflectivity $\eta$ in front of an ideal detector
\cite{Yurke, Kiss}. When $\eta=1$ we have
a perfect detector and when $\eta=0$ the detector no longer detects

\begin{figure}[ht!]
\begin{center}
\includegraphics[scale=0.22]{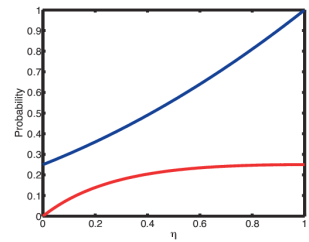}
\end{center}
\caption{\footnotesize Shows the probability of measuring a single photon with an inefficient detector (red) and the probability of observing a single photon in the output given we measure a single photon (blue). Here $r=r_{\text{max}}$. }
\label{PoorMeasBoth}
\end{figure}

We calculate the probability of measuring a single photon at the ideal
detector by taking into account that there may be more than 1 photon present, so many terms will give rise to a single photon event on our ideal detector, each with some probability. This gives the probability of measuring 1 click on our detector and having the state $\ket{n}$ output as
$\text{P}_{n}=n\eta\bigl(1-\eta\bigr)^{n-1}\frac{\tanh^{2n}(r)}{\cosh^{2}(r)}$.
The total probability $\text{P}^d_1$ of our inefficient detector to click for a  single
photon is given by
\begin{eqnarray}
\label{Prob1clickOnly}
\text{P}^d_1=\sum_{n=1}^{\infty}\text{P}_{n}
=\frac{4\eta \sinh^{2}(r)}{\bigl(2-\eta+\eta\cosh(2r)\bigr)^{2}}
\end{eqnarray}
This is shown as the red line in figure (\ref{PoorMeasBoth}). In such a case the probability of having a single photon in the output becomes
\begin{equation}
\text{P}^p_1= \frac{(2-\eta+\eta\cosh 2r)^2}{4\cosh^4r}
\end{equation}
shown as the blue line in figure (\ref{PoorMeasBoth}). 

\begin{figure}[ht]
{\bf (a)}\hskip3.6cm{\bf (b)}
\includegraphics[width=3.5cm,height=2.5cm]{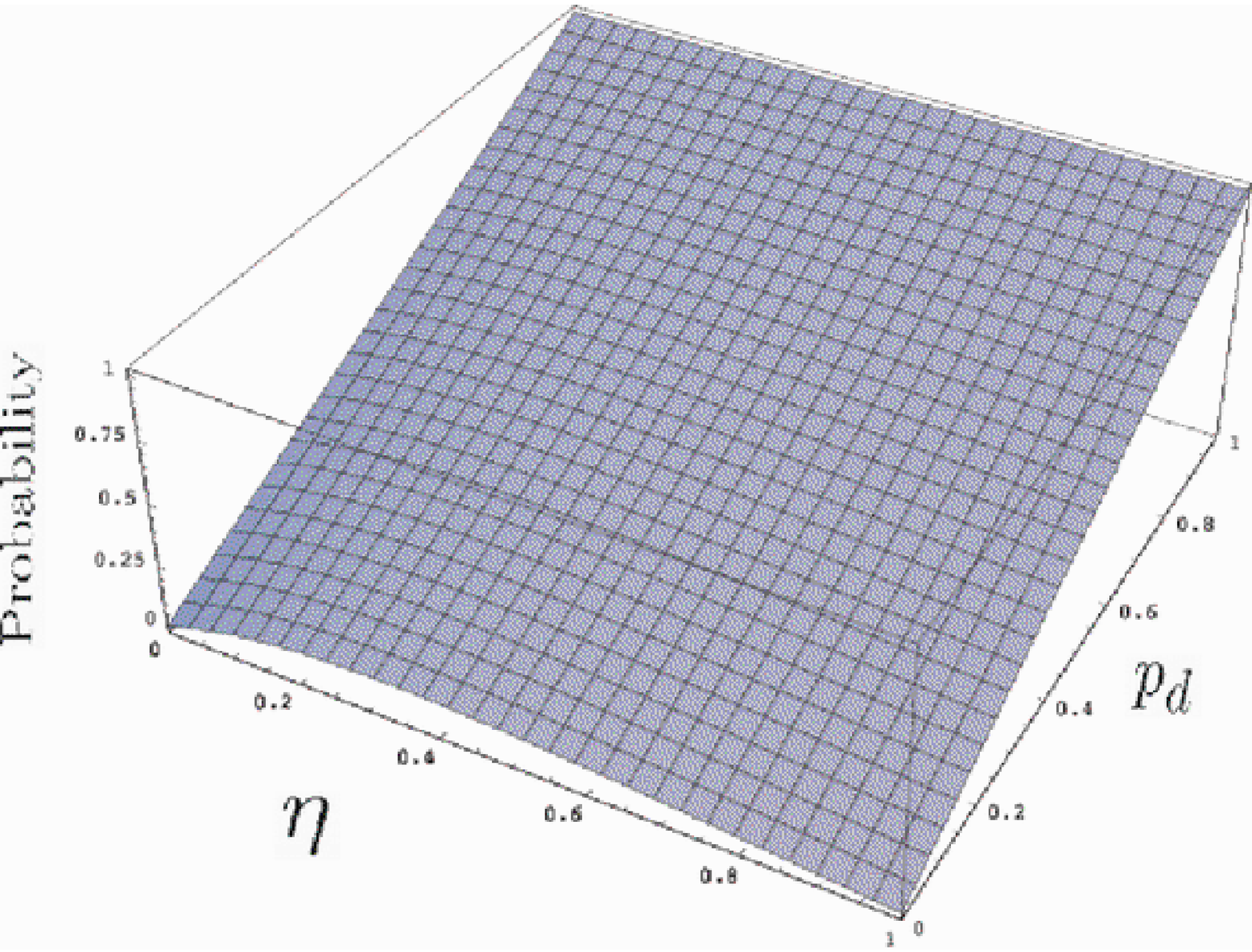}\includegraphics[width=3.5cm,height=2.5cm]{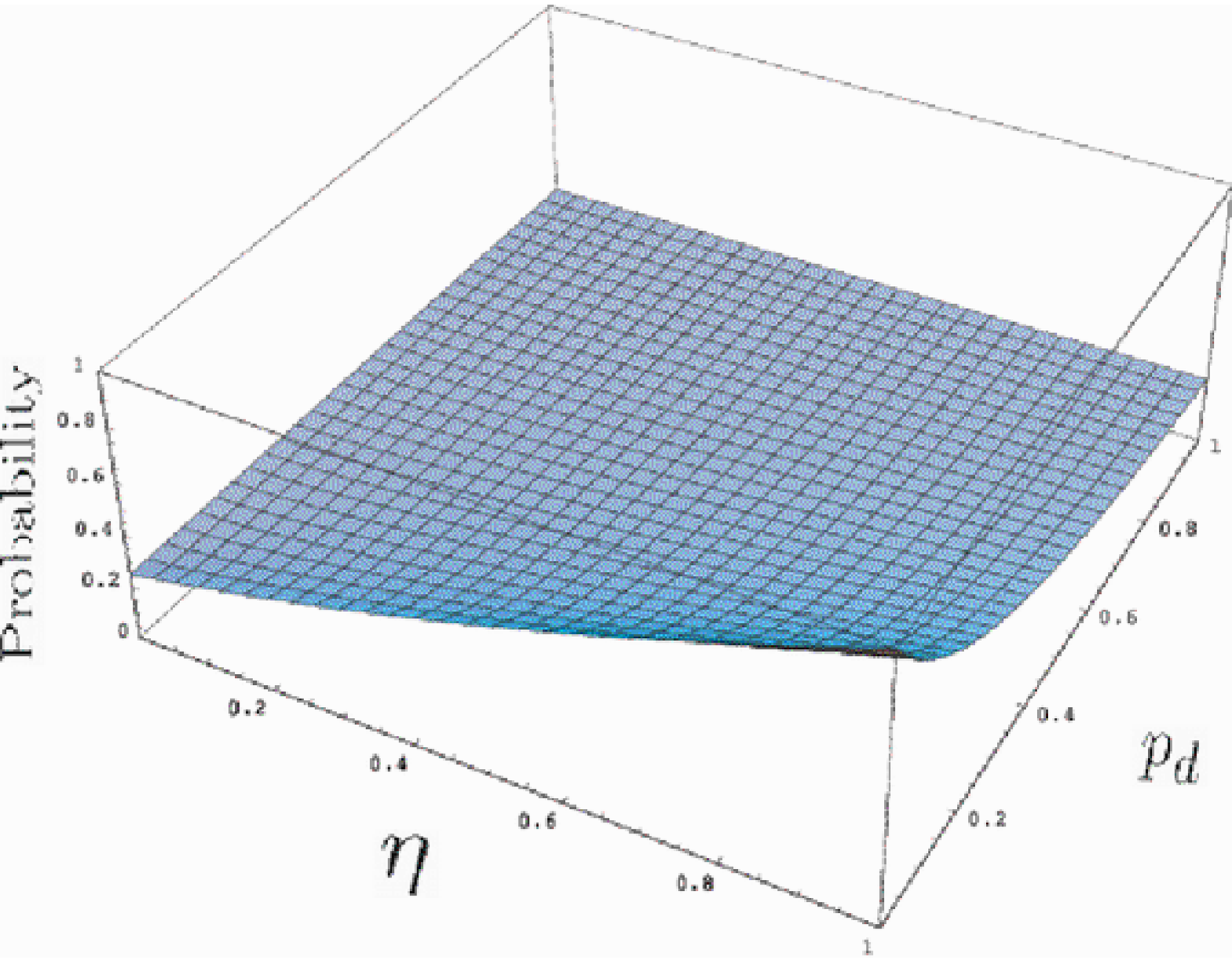}
\caption{\footnotesize {\bf (a)}:  Shows the probability of observing a detector click when we take dark counts into account for our inefficient detector. {\bf (b)}:  Conditional probability of a single photon in the output. Here $r=r_{\text{max}}$. } \label{ProbDarkCountFig}
\end{figure}

We now consider dark counts for the inefficient detector in figure
(\ref{1BSsysFig}). From \cite{Banaszek} it can be shown that the
probability for 1 click when we take dark counts into account is given
by $\text{P}_{1}^{\text{dark}}=p_{d}+(1-p_{d})\langle \hat{P}_{1}\rangle=\langle
p_{d}\cdot\id+(1-p_{d})\hat{P}_{1}\rangle$, where $\hat{P}_{1}=\sum_{n=1}^{\infty}P_{n}\kb{n}{n}$ and $p_{d}$ is the probability of a dark count. Using the
fact that $\id=\sum_{n=0}^{\infty}\kb{n}{n}$ we find the total probability to detect a photon to be 
\begin{eqnarray}
\label{KonradDark2}
p_{d}+\frac{4(p_{d}-1)}{(2-\eta+\eta \cosh2r)^{2}}-\frac{2(p_{d}-1)}{2-\eta+\eta\cosh 2r}
\end{eqnarray}
and the probability to have a single photon output conditioned on detecting a single photon is 
\begin{eqnarray}
\label{KonradDark3}
\frac{(p_{d}(\eta-1)-\eta)(2-\eta+\eta\cosh2r)^{2}\text{sech}^{2} r\tanh^{2} r}{p_{d}(4+(\eta-2)\eta)-2\eta+\eta\cosh 2r (2-2p_{d}(\eta-1)+p_{d}\eta\cosh2r)}
\end{eqnarray}
We plot the total probability (Eqn.(\ref{KonradDark2})) in figure (\ref{ProbDarkCountFig}a) and the conditioned single photon output probability (Eqn.(\ref{KonradDark3})) in figure (\ref{ProbDarkCountFig}b). 

\begin{figure}[ht]
{\bf (a)}\hskip3.6cm{\bf (b)}
\includegraphics[width=3.5cm,height=2.5cm]{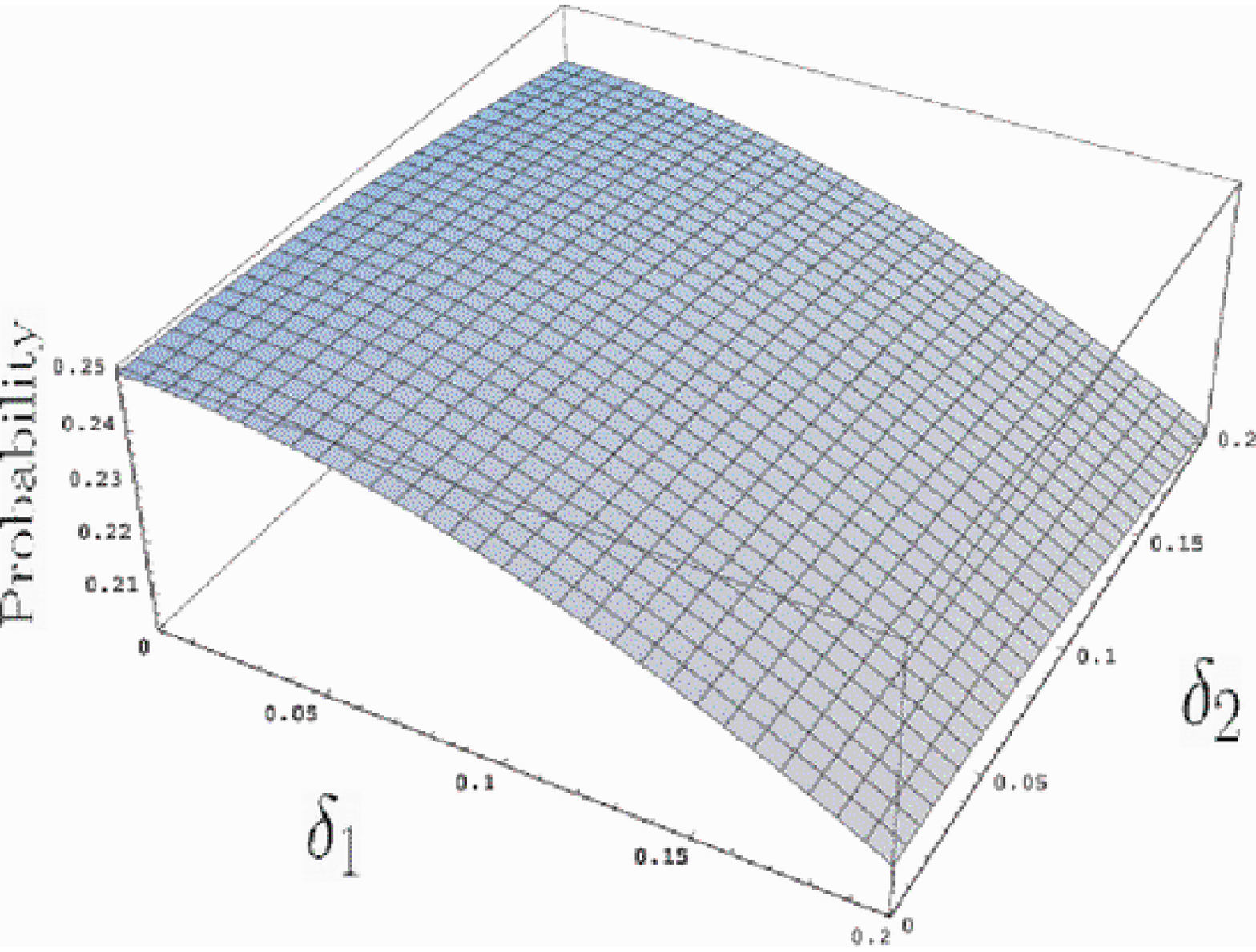}\includegraphics[width=3.5cm,height=2.5cm]{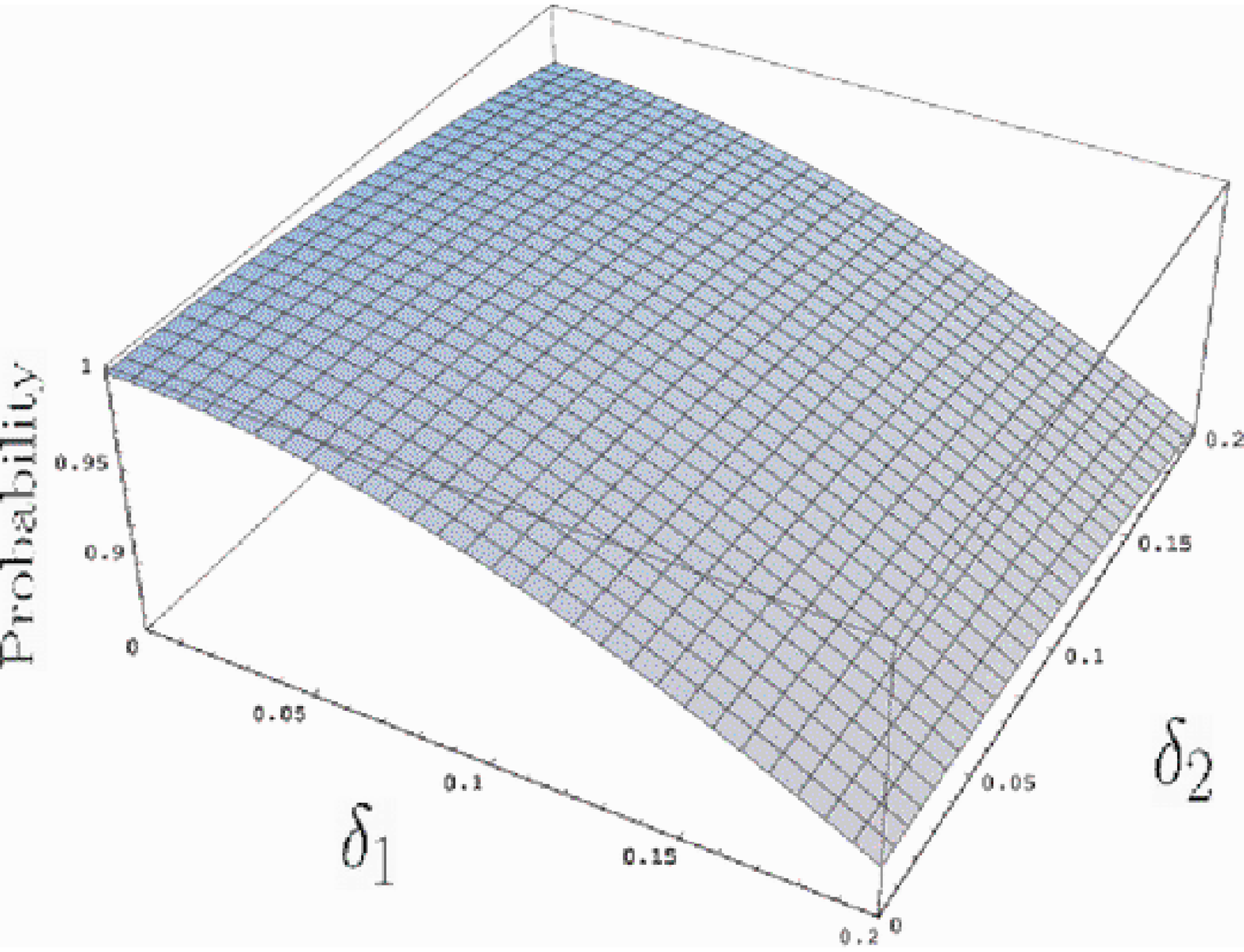}
\caption{\footnotesize {\bf (a)}:  Shows a probability of observing $\ket{1}$ in mode 1 as a function of the beam splitter perturbation $\delta_{1}$ and $\delta_{2}$ {\bf (b)}: Shows a probability of observing $\ket{3}$ in mode 1. Here $r=r_{\text{max}}$.}\label{BadBSAlll3}
\end{figure}

Next we consider a beam splitter that has some uncertainty in the reflectivity. Instead of the usual beam splitter matrix (\ref{Lambda}) we consider the case when the angles $\theta$ and $\phi$ have perturbations given
by: $\theta\mapsto \theta+\delta_{1}$, $\phi\mapsto \phi +\delta_{2}$. When we condition off observing a single photon in the detector the state in the output mode
becomes:
$\ket{\psi}_{\text{badBS}}=
\mathcal{A}_{1}\hat{a}_{1}^{\dagger}\exp[\mathcal{B}_{1}(\hat{a}_{1}^{\dagger})^{2}]\ket{0}$
where $\mathcal{A}_{1}= i e^{(i \varphi)}
\frac{\tanh(r)}{2\cosh(r)}\Bigl(
\cos^{2}\delta_{1}-\sin^{2}\delta_{1}\Bigr)\Bigl( e^{(-i
  \delta_{2})}+e^{(i\delta_{2})} \Bigr)$ and $\mathcal{B}_{1}=
-\frac{1}{4}e^{(i \varphi)} \tanh(r)\Bigl(
\bigl(\cos\delta_{1}-\sin\delta_{1}\bigr)^{2}-e^{(2i\delta_{2})}\bigl(
\cos\delta_{1}+\sin\delta_{1}\bigr)^{2}\Bigr)$.
We see that the probability of having a single photon on the condition
of a click in the detector is
\begin{eqnarray}
\label{badBS1phot}
\text{P}^{\text{BS}}_1=\frac{1}{\sum_{n=0 }^{\infty} \frac{(2n+1)!}{(n!)^{2}}
 || (\mathcal{B}_{1})^{n}||^{2}}
\end{eqnarray}
We see that the probability of observing $n$ photons in mode 1 is
given by 
\begin{eqnarray}
\label{badBSgen1}
\frac{(2n+1)!}{(n!)^{2}}|| \mathcal{A}_{1} (\mathcal{B}_{1})^{n}||^{2}
\end{eqnarray}
The probability of a single photon state output in mode 1 as a function of $\delta_1$ and $\delta_2$
is graphed in the figure (\ref{BadBSAlll3}). In figure (\ref{BadBSAlll3}a) we show the probability of observing 1 photon in mode 1 (Eqn.(\ref{badBSgen1})) and in figure (\ref{BadBSAlll3}b) we show the probability of having a single photon conditioned on a single photon detection (Eqn.(\ref{badBS1phot})). 

In conclusion, we have proposed a single photon source using pulsed degenerate squeezed vacuum states, linear optics and conditional measurements. We have shown that we can in theory get a
certified single photon at every four attempts, with today's achieved value of pulse squeezing this reduces to ten attempts. We have also shown that this process is robust under various imperfections.


We thank M. Knill, A. Imamoglu, B.C. Sanders, D. Berry, G.J. Milburn and K. Banaszek for useful
interaction. CRM and RL are supported in part by CIAR, NSERC, RL is also supported in part by ORDCF and the Canada Research Chair program and ME would like to thank the Swedish Research Council for financial support and also Bj\"orn Hessmo for useful discussions. RL and ME thank the Newton Institute.


\end{document}